\documentstyle[sprocl,epsfig]{article}

\def\J#1#2#3#4{{#1} {\bf #2}, #3 (#4)}

\def\NPB{{\em Nucl. Phys.} B}
\def\NPA{{\em Nucl. Phys.} A}
\def\PLB{{\em Phys. Lett.}  B}
\def\PRL{\em Phys. Rev. Lett.}

\def\PRC{{\em Phys. Rev.} C}

\def\PRPTS{{\em Physics Reports}}

\def\M1bar{\overline{\rm M1}}

\def\be{\begin{equation}}
\def\ee{\end{equation}}
\def\bea{\begin{eqnarray}}
\def\eea{\end{eqnarray}}
\def\roughly#1{\mathrel{\raise.3ex\hbox{$#1$\kern-.75em%
\lower1ex\hbox{$\sim$}}}}

\def\fm{{\rm fm}}
\def\lsim{\roughly<}

\def\vP{{\vec P}}

\def\vs{{\vec \sigma}}

\def\hatk{{\hat k}}
\def\MS{{\mbox{M1}({}^1S_0)}}
\def\MV{{\mbox{M1}({}^3S_1)}}
\def\EV{{\mbox{E2}({}^3S_1)}}
\def\calO{{\cal O}}
\def\calM{{\cal M}}
\def\calB{{\cal B}}
\def\ve{{\vec \epsilon}}

\def\abs#1{{\left| #1 \right|}}
\renewcommand{\thefootnote}{\fnsymbol{footnote}}
\def\rM{{\cal R}_{\rm M1}}\def\rE{{\cal R}_{\rm E2}}
\begin{document}

\hfill {\bf KIAS-P99024, TRI-PP-99-05}
\vskip 0.3cm

\title{ON MAKING PREDICTIONS WITH EFFECTIVE FIELD THEORIES IN NUCLEAR PHYSICS
 \footnote
{Talk given by MR at the Workshop on ``Nuclear Physics with
Effective Field Theories," Institute for Nuclear Theory, University
of Washington, February 25-26, 1999, to appear in World
Scientific.}}

\author{Tae-Sun Park}

\address{Theory Group, TRIUMF, \\
   Vancouver, B.C., Canada V6T 2A3\\
   E-mail: park@alpha02.triumf.ca}

\author{Kuniharu Kubodera}

\address{Department of Physics and Astronomy, University of
South Carolina\\
Columbia, SC 29208, USA\\E-mail: kubodera@nuc003.psc.sc.edu}

\author{Dong-Pil Min}

\address{Department of Physics, Seoul National University,\\
Seoul 151-742, Korea\\E-mail: dpmin@phya.snu.ac.kr}

\author{Mannque Rho}

\address{Service de Physique Th\'eorique, CE Saclay\\
91191 Gif-sur-Yvette, France\\ and \\
School of Physics, Korea Institute for Advanced Study\\ Seoul 130-012, Korea
\\E-mail: rho@spht.saclay.cea.fr}


\maketitle\abstracts{ Based on the effective field theory previously formulated
by us to accurately {\it postdict} all low-energy two-nucleon properties as well
as {\it predict} certain electroweak transitions in heavy nuclei,
we make {\it parameter-free predictions} for the polarized $np$ capture process
$\vec{n}+\vec{p}\rightarrow d +\gamma$ presently being measured at the
Institut Laue-Langevin in Grenoble. Other participants of this meeting
are invited to make their own predictions using their preferred approaches
and join the bet for the best prediction to confront the forthcoming
experiment.}

\setcounter{footnote}{0}
\renewcommand{\thefootnote}{\roman{footnote}}

\section{Introduction}\label{intro}
\indent\indent
This talk was initially meant to be given by Tae-Sun Park since he is
the one doing most of the work -- with a little help from three of
us -- but the organizers asked me (MR) to present
the paper instead. In giving this talk, I would like to distinguish between the
statements I am making for which the other members of the collaboration
should not be held responsible and those that are endorsed by all of us.
The former will be addressed (mostly in footnotes) by ``I" and the latter by
``We."

Since the early attempt to apply effective field theories to nuclear
physics~\cite{weinberg,pmr,vankolck}, there have been many papers written
on the subject, the most recent development of which is being summarized in
this meeting~\cite{thismeeting}.
In confronting Nature with effective field theories in
nuclei
-- one of the main themes of this meeting, one has had to be content mostly
with {\it postdictions},
 genuine {\it predictions} being harder to come by. 
The reason is simply that effective
field theories involve, at each order of power counting, a certain number
of parameters in the effective Lagrangian. It is believed
that those parameters are in
principle calculable from first principles for a given scale ({\it e.g.},
lattice QCD~\cite{lepage}) but in
practice, they have to be fixed by experimental data.
Once the parameters are so fixed, the Lagrangian can then be used to
make predictions for
other processes that involve the same parameters. Up to date, however,
most of the calculations involved fixing of the parameters and only rarely
could one predict and check the prediction by experiments.

In this talk, we would like to present a genuine prediction based on the
formulation of an effective field theory that we have developed
during the last few years. Since our formulation
is available in the literature, we will not dwell on the details of the
formalism but present the essence of the arguments while stressing the
possible caveats involved.
We illustrate how
well postdictions can be made and then how to make predictions for
two specific processes, one for
which data are available and hence the theory can be tested immediately and the
other for which data are not yet available but will be forthcoming, offering
a marvelous possibility for an honest prediction totally unbiased by
available experiments.
\section{The Chiral Filter}\label{chiralfilter}
\indent\indent
An early attempt predating the advent of QCD and effective field
theories in nuclear physics was made in 1978 under a conjecture
called the ``chiral filter hypothesis"\cite{KDR}. Based on current
algebras, it was argued that corrections to the single-particle
transitions in nuclei for the isovector M1 operator and the weak
axial charge operator should be dominated by one-soft-pion-exchange
two-body currents. As a corollary, if the soft-pion exchange
two-body currents are suppressed either by symmetry or kinematics
as for instance in the isoscalar EM current or the Gamow-Teller
operator ({\it i.e.}, the space component of the axial current), the
chiral filter says that corrections to the single-particle operator
cannot be given by a few controlled terms, thereby making the
calculations highly model-dependent. This hypothesis that appeared
to be somewhat ad hoc at the time it was proposed turned out to be
justified in the context of chiral perturbation theory~\cite{pmr}:
The soft-pion exchange term is indeed the leading order correction
in the chiral counting wth the next-to-leading correction
calculable but strongly suppressed.

There are two consequences of this chiral filter that survive in the context of
modern effective field theories. One is a nontrivial
postdiction and the other is an interesting prediction that has been largely
confirmed.
\subsection{A strategy for effective field theory (EFT)}\label{strategy}
\indent\indent
As it stands, there are effectively two ``alternative" ways of
power counting in setting up effective field theories for
two-nucleon systems. One is the original Weinberg
scheme~\cite{weinberg} in which the leading four-Fermi contact
interaction and a pion-exchange are treated on the same footing in
calculating the ``irreducible" graphs for a potential that is to be
iterated to all orders in the ``reducible" channel. The power
counting is done only for the irreducible vertex. The other is the
``power divergence subtraction" (PDS) scheme~\cite{KSW} in which
only the leading (nonderivative) four-Fermi contact interaction is
iterated to all orders with the higher-order contact interactions
and the pion exchange treated perturbatively. While the PDS scheme
is perhaps more systematic in the power counting, we believe that the
Weinberg scheme is not only consistent with the strategy of EFT but
also, in the sense developed below,
 more flexible and predictive with possible errors committed due
to potential inconsistency in the power counting generically suppressed.
In our work, this scheme is adopted.

Iterating to all orders in the reducible channel with the
irreducible vertex is equivalent to solving the Schr\"odinger
equation with a corresponding potential. This then suggests that we use,
in calculating response functions to slowly varying electroweak
fields that we are interested in, those wave functions computed
with so-called ``realistic potentials," the prime example being the
Argonne $v_{18}$ potential~\cite{argonne} (called in short Av18). In fact this
procedure of mapping effective field theory to realistic wave
functions  -- a hybrid approach~\cite{KK} -- for two-nucleon response
functions employed previously by
us~\cite{pmr} has recently been
justified by means of a cutoff regularization~\cite{PKMR}. van Kolck has
presented a similar argument in support of such a hybrid
procedure~\cite{vk}. Now in this framework, the power counting
reduces simply to a chiral counting in the irreducible vertex for
the current as the current appears only once in the graphs. We are
then allowed to separate the current matrix element of interest
into the single-particle and exchange-current terms, with the
single-particle matrix element given entirely by that with the Av18
wave function and the exchange current contribution -- given by the
matrix element with the same wave function -- from operators
computed in standard baryon chiral perturbation theory. The old
soft-pion exchange term figuring in the chiral filter -- if not
suppressed -- is just the leading order contribution in this
series.
\subsection{Unpolarized $np$ capture}
\indent\indent
A good example of postdictions that follow from the above scheme
and that will be a basis for a genuine prediction described below
is the process
\bea
n+p\rightarrow d+\gamma\label{np}
\eea
where both nucleons are unpolarized and the incoming neutron has a
thermal energy~\footnote{The relative momentum in the center of
mass is $\sim 3.4\times 10^{-3}\ \ {\rm MeV}$.}. This process has
been computed to the next-to-next-to-leading order (NNLO) in the chiral
counting in the scheme described above for the current~\cite{pmr2}.
The theoretical cross section $\sigma_{th}=334\pm 3\ \ {\rm mb}$
which agrees with the experiment within the error bar consists of
the leading contribution, $305.6\ \ {\rm mb}$, coming from the
single-particle matrix element given by the Av18 wave function and
the remainder from the exchange current dominated by the soft
one-pion exchange according to the chiral filter.

Two points are worth noting in this result. First, this is a
bona-fide calculation and {\it not} a fit: Within the scheme
adopted here, there are no free parameters.
Secondly as shown in \cite{PKMR}, the single-particle matrix
element has a negligible uncertainty, so that the error in the theory
is entirely attributed to the uncertainty in the exchange-current
operator associated with the short-distance part of the
interactions that cannot be accessed  by chiral perturbation
theory. This part introduces a scale and renormalization-scheme
dependence and only those results that are not
sensitive to this short-distance uncertainty can be trusted. In the
framework in which the realistic wave functions figure, the
short-distance scale is set by the cutoff proportional to
$r_C^{-1}$ where $r_C$ is the ``hard-core radius" that removes the
part of the wave function for $r\lsim r_c\neq 0$. The net effect of
this cutoff is that in addition to cutting the radial integrals in
the coordinate space, it removes {\it all} zero-range terms in the
current operator including zero-ranged counter terms. The $1 \%$
error bar assigned to the theory for (\ref{np})
represents the uncertainty in this
cutoff procedure. This procedure can be justified for the process
in question by using a cutoff $\sim r_C^{-1}$ and showing that the
counter term ``killed" by the hard core is in that uncertainty
range. Given that the four-Fermi counter term is removed by the
hard core which may be viewed as exploiting a scheme dependence,
there are no more parameters in the theory~\footnote{See appendix B
in the second reference of \cite{pmr2} for the zero-ranged counter
term in question. In the framework of \cite{KSW} which avoids
scheme dependence at the expense of predictiveness, this counter
term is a non-negligible parameter, rendering a bona-fide
calculation infeasible~\cite{npKSW}.}. This procedure, familiar
to nuclear physicists, will be referred to as
``hard-core cutoff scheme" (or HCCS in short).
Below we will see that when
the chiral filter does not apply, the removal of the zero-range
counter terms by the simple hard-core cutoff may be suspect.
\subsection{Prediction for the axial-charge transitions in nuclei}\label{axial}
\indent\indent
The EFT scheme described in section \ref{strategy} makes a rather
clean prediction which has not yet been adequately appreciated
in nuclear physics community. The chiral filter idea~\cite{KDR},
by now validated in chiral perturbation theory~\cite{pmr},
predicts that in the nonrelativistic regime, the axial charge
transition matrix element for the $\beta$ decay process in nuclei
\bea
A (J^{\pm})\rightarrow A^\prime (J^\mp) +e +\nu; \ \ \ \Delta
T=1\label{axialcharge}
\eea
should receive a huge one-soft-pion exchange current correction,
amounting to more than 50 \% of the single-particle matrix element.
The prediction for this is quite robust and firm with very little
nuclear model dependence: possible corrections from higher chiral
order terms -- which are calculable a priori -- are estimated to be
less than 10 \% of the leading soft-pion terms~\cite{pmr,pkt}.

How do we go about checking this prediction against experiments?

It is not possible to test this in the two-nucleon systems for
which the effective field theory is most extensively developed: the
matrix element for (\ref{axialcharge}) cannot be measured in
few-nucleon systems. The measurement can only be made in heavier
nuclei. This means that to compare with experiments, we would have
to account for the effect of density or multi-nucleon interactions
on the transition. This introduces a subtlety which is interesting
in its own right.  We will show below that there is  rather
compelling evidence that both the chiral filter and density effect
are confirmed~\footnote{The issue of effective field theory for
hadrons immersed in a hadronic medium with baryon density $\rho$
was one I (MR) would have liked to discuss in this meeting but it
is out of the scope of this presentation. For a recent discussion
on this subject, we simply refer to the article~\cite{frs}. For the
present problem, the net effect is the presence of what is known as
``Brown-Rho (BR) scaling" $\Phi (\rho)$~\cite{BR}.}.

Experimentalists extract the axial charge matrix element ${\cal M}$
from their experiments and then write~\cite{warburton}
\bea
\epsilon_{MEC}={\cal M}_{measured}/{ M}_{th-1b}
\eea
where the numerator is the total ``measured" value or more
precisely the value extracted from experiments and the denominator
is the {\it theoretical} single-particle matrix element of the
single-particle current whose constants are {\it unrenormalized} by
medium.  The model dependence of the denominator -- and to some
extent the numerator -- makes this quantity not entirely empirical,
thus open to controversy among theorists: It would seem that the
``experimental'' $\epsilon_{MEC}$ would in practice depend upon the
model for the wave function used for the sing-particle matrix
element. On the other hand, our claim is that the theoretical
expression for this quantity is well-defined and free of model
dependence~\footnote{ Several people in the audience voiced doubt
as to whether the experimental test I discussed is truly valid. One
of the reasons given is the model dependence of the so-called
empirical information in $\epsilon_{MEC}$. This objection is to
some extent valid and needs to be carefully examined. The presently
employed procedure is to calculate the single-particle matrix
element with the ``best" nuclear wave functions of the
single-particle axial charge operator whose coupling constants are
{\it unrenormalized} by medium. The question then is what does one
mean by
``best" wave function? Here we should stress that
in the case in question, there is a reasonably satisfactory answer.
Let's take Warburton's analyses~\cite{warburton}. Although his analyses do
involve shell-model wavefunctions, 
the semi-empirical nature of the effective transition operator method he 
adopted gives his results much more robustness than people normally expect
from shell-model calculations.  That is, after including the core 
polarization effects in the form of rescaling the single-particle
matrix elements, there is in fact very little room for changing 
nuclear physics input in his analyses.  As a measure of the basic
soundness of his approach, we mention the following fact.  As far as 
nuclear dynamics is concerned, the rank-zero and rank-one first-forbidden 
transitions share the same feature, but the exchange currents for them
have very different behavior.  The rank-one operators coming from the 
space component of the axial current has only small exchange currents
whereas the time component of the axial current that contributes to
the rank-zero operator is expected to have a {\it huge} exchange current.  
Now, Warburton's calculations reproduce very well the strengths of all
the rank-one matrix elements with no further adjustments whereas,
for the rank-zero matrix elements, he found it clearly necessary to
introduce the ``empirical'' enhancement factor $\epsilon_{MEC}$.   
   
The other reason (which is theoretical) is
that the BR scaling is so far implemented in the limit of infinite
nuclear matter and the question arises regarding the finite size
effect of the nuclei measured. My answer to this objection is that
whereas experimentally the quantity $\epsilon_{MEC}$
may perhaps be somewhat model-dependent, theoretically, however,  it is very
well defined: It involves the BR scaling factor $\Phi$ which is
smoothly varying for density up to $\rho=\rho_0$, so what matters
is the average density involved and the quantity ${\cal R}$, the
ratio of the matrix element of the exchange current over that of
the single-particle operator, which is quite insensitive to nuclear
models provided the same wave functions are used for both.}.

Let ${\cal M}_{total}$ denote the total theoretical axial-charge matrix
element and ${\cal M}_{n}$ with $n=1, 2$ be the matrix element of the $n$-body
operator effective in medium. Then the prediction is that
\bea
{\epsilon_{MEC}}^{th}&=&{\cal M}_{total}/M_1=\Phi^{-1} (1+{\cal R}),
\label{eth}\\
{\cal M}_{total}&=&{\cal M}_1 +{\cal M}_2
\eea
with
\bea
{\cal R}={\cal M}_2/{\cal M}_1= \left(M_2/M_1\right) (1+{\cal O} (10^{-1}))
\eea
where $M_n$ is the matrix element of the $n$-body current with its
basic coupling constants ({\it e.g.}, $f_\pi$, $g_A$ etc) {\it
unrenormalized} by medium. This relation follows within our HCCS
(hard-core cutoff scheme) provided one assumes that all light-quark
hadron mass $M$ {\it except for the pion mass} scales in medium as
\bea
M^\star/M\approx f_\pi^\star/f_\pi\approx \Phi (\rho).
\eea
The expression (\ref{eth}) was first derived by Kubodera and
Rho~\cite{KR91} and later corrected~\cite{pkt,frs} \footnote{There
was an error in the original derivation due to the fact that the
non-scaling of the pion mass, indicated by both theory and
experiment, was not properly taken into account.}. Now what we need
is the value for $\Phi$ and ${\cal R}$ for a range of nuclei that
have been measured. The range of nuclear density involved is $1/2
\lsim \rho/\rho_0 \lsim 1$ where $\rho_0$ is the nuclear matter
density. For this range, we know from information gotten from giant
dipole resonances and QCD sum rules in medium~\cite{frs} that
$\Phi$ goes as $\sim 1/(1+0.28 (\rho/\rho_0))$,
\bea
0.78\lsim \Phi\lsim 0.88 \ \ {\rm for}\ \ 1/2\lsim \rho/\rho_0\lsim 1.
\eea
The two-body current in the ratio ${\cal R}$ is given within 10\% by a
soft-pion
exchange which is completely fixed by chiral symmetry. The ratio turns out to be
extremely {\it insensitive} to nuclear models used to
compute the matrix elements
and depends only slightly on density. It comes out to
be \footnote{Owing to the chiral filter applicable to
this process, the uncertainty in our hard-core cutoff scheme, {\it i.e.},
``killing'' zero-range counter terms, 
is reflected in the 10\% uncertainty in ${\cal R}$
mentioned above.}
\bea
0.43\lsim {\cal R}\lsim 0.61 \ \ {\rm for}\ \ 1/2\lsim \rho/\rho_0\lsim 1.
\eea
This gives the range
\bea
1.63\lsim \epsilon_{MEC}^{th}\lsim 2.06 \ \ {\rm for}\ \ 1/2\lsim\rho/\rho_0
\lsim 1.
\eea
A formula as simple as (\ref{eth}) must have a simple and clean test.
It must be readily confirmed or infirmed.

We claim that there is a strong empirical support
for this prediction. Indeed
this prediction can be compared with the presently available
``empirical" values~\cite{minamisono,baumann,vangeert,warburton}
\bea
\epsilon_{MEC}^{exp}&=& 1.64\pm 0.05\
\ (A=12)  \
; 1.62\pm 0.05\ \ (A=50)  \ ;\nonumber\\
&& 1.95\pm 0.05\ \ (A=205) \ ; 2.05\pm 0.05\ \ (A=205\sim 212).
\eea
We further suggest that these constitute evidence for {\it both}
(1) the gigantic enhancement predicted by the chiral filter,
ranging from 46\% to 61 \% and (2) the additional enhancement
predicted by BR scaling, ranging from 20 \% to 45 \%. There is of
course the caveat that individual transitions must be subject to
some finite-size effects but the point is that what one is probing
here is a generic bulk property of nuclear matter -- which to us is
the most interesting part of the story.
\section{Polarized Neutron-Proton Capture: A New Probe}
\indent\indent
We will now go outside of those processes protected by the chiral filter
and make a genuine prediction even though the process,
unprotected by the chiral filter, could be highly suppressed.
One would think that such a prediction is out of the scope of
effective field theories but surprisingly, it turns out not to be the case
in our EFT scheme.
\subsection{Selection rules}
\indent\indent
Consider the process
\bea
\vec{n}+\vec{p}\rightarrow d+\gamma\label{polnp}
\eea
where now both the target proton and the projectile neutron are
polarized. This process is being measured at the Institut
Laue-Langevin (ILL) in Grenoble by M\"uller {\it et al}.~\cite{mueller}
The interest in this experiment is that with the polarized target
and the polarized beam, one can measure small matrix elements that
are overwhelmed by the dominant isovector M1 matrix element when
averaged over polarization. To see this, look at the quantum
numbers involved in the process. The initial state of (\ref{polnp})
at very low energy we are dealing with can be in either $^1S_0$ or
$^3S_1$ channel. The Fermi-Dirac statistics requires that the
former must be in $T=1$ and the latter in $T=0$ where $T$ is the
isospin. The final nuclear state in (\ref{polnp}) is the deuteron
which is in $^3S_1$ or $^3D_1$ with $T=0$. There are then three
relevant transition matrix elements with the emission of a soft
photon, {\it i.e.}, isovector M1, isoscalar M1 (which we shall denote
from now on as $\M1bar$ to distinguish it from the isovector M1)
and isoscalar E2. We shall adopt the convention of M\"uller et
al~\cite{mueller} and write the transition amplitude as
\bea
\langle \psi_d(M_d), \gamma(\hatk \lambda) | {\cal T}|
    \psi_{np}(s_p, s_n)\rangle
 = \chi^\dagger_{1 M_d}\, \calM(\hatk, \lambda)\, \chi_{s_p} \chi_{s_n}
\eea
with
\bea
\calM(\hatk \lambda) =
 \sqrt{4\pi} \frac{\sqrt{v_n}}{2 \sqrt{\omega} A_s}
 \, \left[
  i (\hatk \times \epsilon^*)\cdot (\vs_1-\vs_2)\, \MS
\right.
\nonumber \\
 \left.
  - i (\hatk \times \epsilon^*)\cdot (\vs_1+\vs_2)\,\frac{\MV}{\sqrt{2}}
  + (\vs_1\cdot\hatk \vs_2\cdot \epsilon^*
   + \vs_2\cdot\hatk \vs_1\cdot \epsilon^*)
    \frac{\EV}{\sqrt{2}}
\right]\label{amp}
\eea
where $M_d$ and $\lambda$ are respectively the polarizations of the deuteron
and the photon, $\hatk$ is the unit momentum vector of the
photon, $\omega$ its energy, $\ve \equiv \ve(\vec k \lambda)$, $v_n$ is the
velocity of the neutron and $A_s$ is the deuteron normalization factor
$A_s\simeq 0.8850\ {\rm fm}^{-1/2}$.
In the way defined, the quantities $\MS$, $\MV$ and $\EV$ all
have a dimension of $\fm$
and the cross section for the unpolarized $np$ system takes the form
\bea
\sigma_{unpol}= \abs{\MS}^2 + \abs{\MV}^2 + \abs{\EV}^2\,.\label{xsection}
\eea
The first term is the isovector M1 contribution, the second
the isoscalar M1 and the last the isoscalar E2. As we shall see
below,  the second and third terms
are strongly suppressed compared
to the first, $\sim O(10^{-6})$, so the unpolarized cross section
cannot ``see'' these terms.
\subsection{Polarization observables}
\indent\indent
In order to see those small isoscalar terms, one measures polarization observables,
{\it i.e.}, the photon circular polarization $P_\gamma$ and the photon anisotropy $\eta$.
For an unpolarized proton and a polarized neutron with the polarization vector
$\vec{P}_n$, $P_\gamma$ is given by
\bea
P_\gamma = \frac{|\vP_n|}{2}\,
 \frac{2 \sqrt{2} (\rM - \rE) + (\rM+\rE)^2}{1 + \rM^2 + \rE^2}\label{pgamma}
\eea
where we have defined the ratios
\bea
{\cal R}_{\rm M1} &\equiv& \frac{\MV}{\MS},
\nonumber \\
{\cal R}_{\rm E2} &\equiv& \frac{\EV}{\MS}.\label{ratio}
\eea
Since these ratios turn out to be $\sim 10^{-3}$, eq.(\ref{pgamma})
simplifies with high accuracy to
\bea
P_\gamma \simeq |\vP_n|\, \sqrt{2} (\rM - \rE).
\eea

The anisotropy measures the fully polarized $np$ system involving the
polarization of the target and projectile nucleons and is given by
\bea
\eta&=& \frac{I(90^\circ) - I(0^\circ)}{
     I(90^\circ) + I(0^\circ)}
\nonumber \\
&=& pP\, \frac{\rM^2 + \rE^2 - 6 \rM \rE}{
 4 (1 - pP) + (4 + pP) (\rM^2 + \rE^2) + 2 pP\, \rM \rE}
\eea
where $I$ is the photon intensity,
\bea
pP\equiv \vP_p\cdot\vP_n\,
\eea
and the angle in $I$ measures the photon direction with respect the spin
polarization of the neutron and the proton. Note that unless the factor
$(1-pP)\sim 0$ the anisotropy $\eta$
will be quadratically suppressed while $P_\gamma$ is linear in the ratio.
If however
$(1-pP)\sim 0$, then the anisotropy could be substantial, supplying
an additional formula that would allow one
to extract the two ratios (\ref{ratio}).
The purpose of the experiment is to determine these two ratios.
The quantity $P_\gamma$
has already been measured before by Bazhenov {\it et al}~\cite{russian}, so the aim
of the ILL experiment~\cite{mueller} is to measure the anisotropy $\eta$.
\subsection{Doing EFT}
\indent\indent
As mentioned in section \ref{chiralfilter}, the current involved
here is not protected by the chiral filter. This means that the
soft-pion exchange which is entirely given by chiral symmetry 
considerations
cannot contribute. The corollary to the chiral filter hypothesis
would then suggest that we might be opening  Pandora's box. To our
surprise, this does not seem to be the case for the problem at
hand.
\vskip 0.3cm

$\bullet$ {\bf Power Counting}
\vskip 0.3cm

Since the isovector M1 operator is calculated~\cite{pmr} very
accurately to ${\cal O} (Q^3)$ relative to the single-particle
operator ({\it i.e.}, M1$(^1S_0)=5.78\pm 0.03$ fm in the notation of
(\ref{xsection}) which comes at ${\calO} (Q^{-2}$))
\footnote{Unless otherwise specified, we will always give counting
relative to the leading single-particle operator.}, we will focus on
the isoscalar operators here.

\begin{figure}[htbp]
\begin{center}
\epsfig{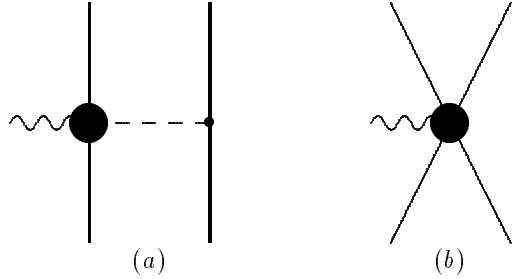}
\caption[gene]{\protect \small
Generic diagrams for the two-body isoscalar current $\calB_{\rm 2B}^\mu$.
The solid circles include
counter-term insertions and (one-particle irreducible)
loop-corrections. The wiggly line stands for the external field
(current) and the dashed line the pion. One-loop corrections for
the pion propagator and the $\pi NN$ vertex are of course to be
included at the same order.}\label{fig1}
\end{center}
\end{figure}

We assume that as in the case studied so far~\cite{PKMR}, given the
accurate wave functions, the leading single-particle matrix elements are
accurately given for both $\M1bar$ ({\it i.e.}, isoscalar M1) 
and E2 operators. The power counting that
we adopt~\cite{weinberg,pmr,PKMR}
then shows that while the ratio
$\langle {\rm 2-body}\rangle/\langle {\rm 1-body}\rangle$
goes like ${\calO} (Q^1)$ for the operator protected by the chiral filter
({\it e.g.}, the isovector M1), the ratio for the isoscalar operators goes
like ${\calO} (Q^n)$
with $n\geq 3$, so naively one would expect a suppression by two orders of
power counting.
Now the question is: How well can we pin down such two-body terms and
higher-order
corrections to them?

\begin{figure}[htbp]
\begin{center}
\epsfig{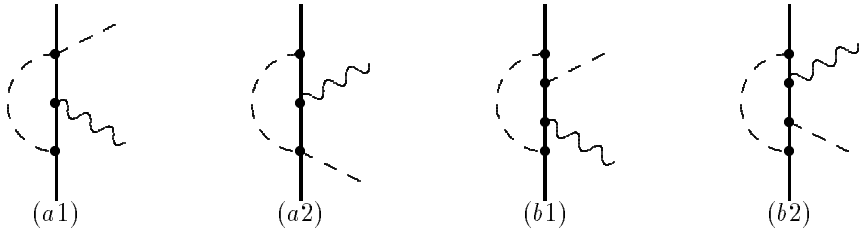}
\caption[deviate]{\protect \small
One-loop graphs that contribute to the
$\calB \pi NN$ vertex where $\calB$ is the isoscalar current.
They give rise to $\calO (Q^4)$ and higher corrections to the leading
order (LO) one-body term.
\label{fig2} }
\end{center}
\end{figure}

\begin{figure}[htbp]
\begin{center}
\epsfig{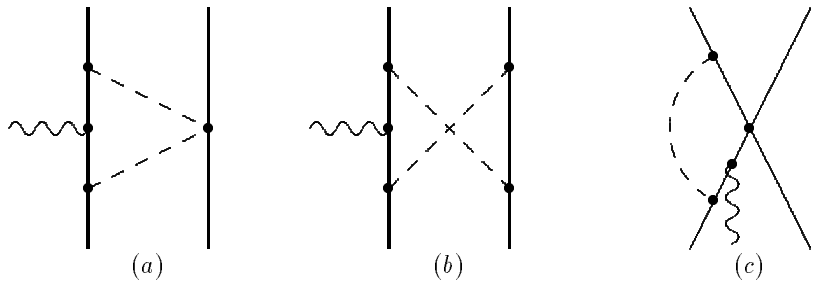}
\caption[deviate]{\protect \small
One-loop graphs that contribute to the two-body baryonic currents.
They come at $\calO (Q^4)$ and higher order relative to the LO
one-body term. All possible insertions of the external line are
understood.
\label{fig3}
}
\end{center}
\end{figure}

In our scheme, there are two classes of irreducible two-body terms in
the current
operator. The class A term consists of graphs with the
one-pion exchange involving an $NN\pi\gamma$ vertex (see Fig.\ref{fig1}a)
and the class B term consists of graphs with two- or more-pion exchanges
(see Fig.\ref{fig1}b).
Since there is no chiral-filter-protected one-pion exchange in the
isoscalar vector current,
the class A term receives
its leading contribution from one-loop corrections (see Fig.\ref{fig2}).
The class B term is generically given by the loop graphs of Fig.\ref{fig3}.
The loop and derivatives (in Fig.\ref{fig3}c)
account for the additional power suppression.

The graphs in Figs.\ref{fig2} and \ref{fig3} typically contribute to the
isoscalar M1 operator ($\M1bar$) and E2 operator at $\calO (Q^4)$.
This is however not the
whole story to $\calO (Q^4)$, for there are so-called
``counter terms'' that can come in at $\calO (Q^3)$. There are in
fact two such terms in the case that we are concerned with.  One
such term is a one-pion exchange graph in Fig.\ref{fig1}a with the
$\calB\pi NN$ vertex given by a finite counter term. This term that
contributes to the $\M1bar$ operator but not to the E2 operator is
dominated by the $\gamma\rho\pi$ coupling in the anomalous parity
component of the effective chiral Lagrangian, {\it i.e.}, the Wess-Zumino
term, which is connected with the Adler-Bell-Jackiw triangle
anomaly. This term is known and so brings in no unknown
parameters. The other is a four-Fermi counter term, call it $g_4$, in
Fig.\ref{fig1}b contributing only to $\M1bar$. The coefficient of
this counter term,
not known a priori, needs to be fixed in the usual way.

Note that there are no $\calO (Q^3)$ counter term
contributions to the E2.

\vskip 0.3cm
$\bullet$ {\bf There is No Parameter in the Theory}
\vskip 0.3cm

We shall now argue that while there remains one parameter
undetermined by the theory it can be gotten rid of in either of the
two ways in which physics at short distance is treated.

We first note that there are no unknown parameters in the class A
diagrams: As mentioned, the $\calO (Q^3)$ counter term is fixed by
the Wess-Zumino term and the loop graphs are completely calculable
to $\calO (Q^4)$ without any parameters. In the class B diagrams,
there are two parameters, call them $V_i$ with $i=1, 2$, associated
with Fig.\ref{fig3}c. One of them, $V_1$, contributes to both
$\M1bar$ and E2 and the other, $V_2$, only to $\M1bar$. We find that the
$V_1$ term plays no role in $\M1bar$ or E2, as it is
suppressed by some power of $Qr_c \ll 1$.

The upshot of all this is that we are finally left with the
four-Fermi counter term $g_4$ and $V_2$. Furthermore, both of them
are associated with zero-range terms in the coordinate space. The
combination that appears here will be called, in short, the $g_4``+"
V_2$ term. So we can combine them with the $\calO (Q^3)$
zero-ranged operator that comes from the one-pion exchange term
with the Wess-Zumino vertex (we shall call this in short  the $\calO
(Q^3)$ WZ term), thereby reducing them effectively to only one
parameter in $\M1bar$. The E2 operator is completely free of
parameters to $\calO (Q^4)$.

Now in the HCCS, the single parameter of the theory does not figure
since the contact operator that is multiplied by the parameter gets
suppressed by the hard core.

But there is a possible caveat here: Since the leading order of
this term is $\calO (Q^3)$ whereas the finite loop corrections are
of $\calO (Q^4)$, it is not obvious that the HCCS should be
reliable in the present case. One might suspect that here
short-distance physics could intervene more strongly than when the
chiral filter is operative. There are two possible remedies to this
problem. One is to implement the operator-product-expansion (OPE)
factorization in the wave function suggested by
Lepage~\cite{lepagelecture} and the other is to use a cutoff $\sim
r_c^{-1}$ in Fourier-transforming the current operators into the
coordinate-space form. We shall use the latter which is a lot
simpler. We use an equivalent method which is to replace the delta
function in all zero-ranged operators by the delta-shell form
\bea
\delta (r)\rightarrow \delta (r-r_c).\label{deltashell}
\eea
This procedure allows the $\calO (Q^3)$ contact operator to
contribute in $\M1bar$, hence allowing to fix the unknown $g_4``+"
V_2$ parameter by fitting the deuteron magnetic moment. It turns
out however that this contact term is dominated by the known $\calO
(Q^3)$ WZ term, so in practice, the unknown parameter plays only a
minor role here. We shall call this scheme the ``modified hard core
cutoff scheme."

\subsection{Our predictions}
\indent\indent
We shall now make {\it our} predictions.
\vskip 0.3cm
$\bullet$ {\bf Hard Core Cutoff Scheme (HCCS)}
\vskip 0.3cm

With the hard core in the wave function, the parameter-dependent
term is killed, so we can now predict the deuteron
magnetic moment $\mu_d$, the quadrupole moment $Q_d$, and the ratios
$\rM$ and $\rE$. There must of course be some dependence on the hard core
size which enters as a cutoff but
the consistency of EFT requires that the cutoff
dependence be small.
\vskip 0.2cm

{\bf Deuteron magnetic moment} $\mu_d$: There is a strong
cancellation between the $\calO (Q^3)$ and $\calO (Q^4)$ two-body
terms, leaving the one-body term essentially uncorrected: The net
two-body correction is found to be less than 0.7\% of the one-body
term. The predicted values for $r_c=0.01, 0.2, 0.4, 0.6, 0.8$ fm
are (in units of nuclear magneton)
\bea
\mu_d = 0.8408,\ 0.8443,\ 0.8426,\ 0.8407,\ 0.8390.
\eea
The experimental value is $\mu_d^{exp}=0.8574$.
\vskip 0.2cm
This small discrepancy will be exploited later to fix the one
parameter that figures when the zero-range operator is not killed
by the hard core as in MHCCS.

{\bf Deuteron quadrupole moment} $Q_d$:
The two-body correction is equally tiny, less
than 0.6\% and more or less independently of $r_c$ within the range
$0.01\lsim r_c\lsim 0.8$ fm. The result is (in unit of fm$^2$)
\bea
Q_d= 0.2710\label{Qd}
\eea
to be compared with the experiment $Q_d^{exp}=0.2859\ {\rm fm}^2$.
It appears that the 5\% discrepancy found here cannot be understood
in low-order effective field theories.
\vskip 0.2cm

$\rM$: The two-body terms of $\calO (Q^3)$ and $\calO
(Q^4)$ in $\M1bar$, separately,
are of the same magnitude as the one-body term, so
although naively suppressed in the power counting, there is no
genuine suppression according to the hierarchy of the order.
 However there is a considerable cancellation between
the two higher-order terms,
leaving the correction to be between 9\% and 25\% of the
single-particle value. The predicted values for $r_c=0.01, 0.2,
0.4, 0.6, 0.8$ fm are
\bea
-\rM\times 10^3=0.869,\ 0.788,\ 0.826,\ 0.871,\ 0.887.
\eea

$\rE$: As in the quadrupole moment, the two-body correction to the
E2 matrix element is small $\lsim 0.4\%$, so the result is
essentially given by the one-body term. Thus within the $r_c$ range
considered, the result is, independently of $r_c$,
\bea
\rE\times 10^{3}= 0.242.\label{RE}
\eea

{\bf Photon circular polarization} $P_\gamma$:
With the above values for $\rM$ and $\rE$,
the predicted values for $r_c=0.01, 0.2,
0.4, 0.6, 0.8$ fm are (for $|\vec{P}_n|=1$)
\bea
-P_\gamma\times 10^3=1.57,\ 1.46,\ 1.51,\ 1.57,\ 1.60.
\eea

{\bf Photon anisotropy} $\eta$: Here the prediction is extremely sensitive to
the value of the polarization $pP$. We will quote for three cases:
$pP=1$ (the ideal case), $pP=0.96$ (the highest polarization that may be reached),
and $pP= (0.5)^2$ (a case most certainly accessible to the experiment).
The predicted values for  $r_c=0.01, 0.2,
0.4, 0.6, 0.8$ fm are
\bea
\eta [pP=1]&=& 0.57, \ 0.61, \ 0.59,\ 0.57, \ 0.56,\nonumber\\
\eta [pP=0.96]\times 10^{5}&=& 1.3, \ 1.1, \ 1.2, \ 1.3, \ 1.3,\nonumber\\
\eta [pP=(0.5)^2]\times 10^7&=& 1.7, \ 1.5, \
1.6, \ 1.7, \ 1.8.
\eea

\vskip 0.3cm
$\bullet$ {\bf Modified Hard Core Cutoff Scheme (MHCCS)}
\vskip 0.3cm

We shall apply the ``smoothing'' (\ref{deltashell}) to the delta
function and account for the term carrying information on
the single parameter
available in the theory. We have the possibility to fix the constant
by ``fine-tuning'' it to the deuteron magnetic moment, that is,
attributing the 5\% discrepancy in $\mu_d$ from the experimental
value to the counter term containing the $\calO (Q^3)$ WZ term and
the $g_4``+" V_2$ term. As mentioned, in practice, this term is
completely dominated by the former (in fitting the deuteron
magnetic moment); therefore the unknown constant plays only a minor
role in the resulting isoscalar M1 operator that is to be used to
compute the $\M1bar$ matrix element.

While $\mu_d$ is no longer predicted in this scheme (since it is
used to pin down the small $g_4``+" V_2$ term), the deuteron quadrupole
mement $Q_d$ remains unmodified from eq.(\ref{Qd}). The ratios
$\rM$ and $\rE$ are of course predicted.
\vskip 0.2cm

$\rM$: There turns out be a remarkable $r_c$ independence for this
quantity. In fact, in the range $0.01\leq r_c\leq 0.8$ fm, the
result is the same:
\bea
-\rM\times 10^3= 0.500.
\eea

$\rE$: This quantity remains unchanged from the HCCS, (\ref{RE}),
\bea
\rE\times 10^3= 0.242.
\eea

$P_\gamma$ and $\eta$: Independently of $r_C$ in the range $0.01\leq r_c \le 0.8$ fm,
we find
\bea
-P_\gamma\times 10^3=1.05
\eea
and
\bea
\eta [pP=1]&=& 0.80,\nonumber\\
\eta [pP=0.96]&=& 0.62\times 10^{-5},\nonumber\\
\eta [pP=(0.5)^2]&=& 0.86\times 10^{-7}.
\eea
\section{Conclusion: Call for a Bet}
\indent\indent
The values we have are not the final ones. First of all,
they will have
to be rechecked more thoroughly, and secondly, given that the isoscalar
matrix elements are so suppressed relative to the isovector matrix
element, it may be necessary to take into account the usually negligible
isospin violation (both in the interaction and electromagnetic radiative
corrections). In any event,
we shall give our preliminary predictions
here with the warning that they are subject to further changes.
We expect to be able to publish a paper with our final numbers in the near
future with the above caveats taken into account~\cite{pkmrILL}.

In Table \ref{prediction} are summarized our predictions. Recall
that there are two schemes for treating the zero-ranged counter
terms. One scheme, referred to as hard core cutoff scheme (HCCS),
is the usual nuclear physics practice to kill the delta function
terms in the operator. The physics so purged from them is
presumably shifted to the finite matrix elements affected by the
``correlation hole."  In this case there are no more parameters
left to account for the small deviation from the experimental data
in $\mu_d$ and $Q_d$ and the results would depend on $r_c$; the
dependence would of course be weak if the procedure were consistent
with the premise of EFT. The other scheme, called modified hard
core cutoff scheme (MHCCS), exploits the non-vanishing of zero-range
operators to fix one parameter available in the theory by fitting
the deuteron magnetic moment, which then determines completely the
isoscalar M1 operator to $\calO (Q^4)$. In this scheme, there is
practically no $r_c$ dependence and the convergence of the
higher-order terms is assured by the procedure (specifically, there
is no difference between the $\calO (Q^3)$ calculation and the
$\calO (Q^4)$ calculation once $\mu_d$ is fit).

\begin{table}[ht]
\begin{center}
\begin{tabular}{|r||r|r|}
\hline
Hard-core scheme &HCCS  &MHCCS
\\ \hline
$10^3\times P_\gamma$ & $-1.5\pm 0.1$ & $-1.1$ \\
\hline
$\eta[pP=1]$ &
$0.58\pm 0.03$ & 0.80 \\
$10^5\times \eta [pP=0.96]$ & $1.2\pm 0.1$ & 0.62 \\
$10^7 \times \eta[pP=0.25]$ &
$1.6\pm 0.1$ & 0.86 \\
\hline \hline
$10^3 \times {\cal R}_{M1}$ & $-0.84\pm 0.05$ & $-0.51$
\\
$10^3 \times {\rE}$ & 0.24& 0.24\\
\hline
\end{tabular}
\end{center}
\caption{\protect \small Predictions using two schemes for implementing
the hard core: HCCS (hard-core cutoff scheme) in which zero-ranged operators
are ``killed'' by the short-range correlation function in the wave function
and MHCCS (modified hard core cutoff scheme) in which the delta function
of zero-range operators is ``smoothed'' to the delta shell form. The
``error bar'' represents the variation over the range $0.01\ {\rm fm}\leq
r_c\leq 0.8\ {\rm fm}$. No error bar means that there is no $r_c$ dependence.
}\label{prediction}
\end{table}
There is only one polarization data available at the moment, namely,
the photon circular polarization measured by the Russian group~\cite{russian}
\bea
P_\gamma^{exp}=(-1.5\pm 0.3)\times 10^{-3}.\label{russian}
\eea
As it stands, this experiment seems to favor the HCCS result.
However the prediction in which we have more confidence is that of
MHCCS, which puts it somewhat lower than the experimental value
(\ref{russian}). Clearly additional measurements will be needed to
confirm (\ref{russian}) or improve on it.

\vskip 0.3cm
$\bullet$ {\bf The Bet}
\vskip 0.3cm

{\it We would now like to invite other workers in the field --
particularly those who are in this audience, who have worked out
consistent and systematic power countings -- to make similar
predictions and participate in the  bet for the  best prediction,
that is, the one that agrees best with the experimental result
that is forthcoming. As for our prediction, the most interesting
possibility would be that the experiment simply disagrees with
our two versions of the hard-core scheme. That would sharpen our 
chiral filter
conjecture and bring in totally new physics.} 
\footnote{ At the time of my talk, I made a
specific proposal for the prize for the winner and I confirm that
proposal here. I would propose that each participant should be
willing to offer to the winner a bottle of one of the {\it best}
wine of his/her country. As for me, I would offer a bottle of
superb French wine, Chateau
Mouton-Rothschild, a ``premier grand cru class\'e.''}
\section*{Acknowledgments}
\indent\indent
We would like to thank the organizers of this meeting for the invitation
to participate in the exciting debate
and Thomas M\"uller for discussions and correspondence on
the on-going experiment at the ILL.

\end{document}